\documentclass[prl, twocolumn, showpacs, superscriptaddress]{revtex4-2}
\usepackage{amsmath, amssymb}
\usepackage{graphicx}
\renewcommand{\section}[1]{{\par\it #1.---}\ignorespaces}

\begin{document}
\title{Exotic Behavior of Parity in the Superradiant Phase of Quantum Rabi Model}
\author{Yun-Tong Yang}
\affiliation{School of Physical Science and Technology, Lanzhou University, Lanzhou 730000, China}
\affiliation{Lanzhou Center for Theoretical Physics $\&$ Key Laboratory of Theoretical Physics of Gansu Province, Lanzhou University, Lanzhou 730000, China}
\author{Junpeng Liu}
\affiliation{School of Physical Science and Technology, Lanzhou University, Lanzhou 730000, China}
\affiliation{Lanzhou Center for Theoretical Physics $\&$ Key Laboratory of Theoretical Physics of Gansu Province, Lanzhou University, Lanzhou 730000, China}
\author{Hong-Gang Luo}
\email{luohg@lzu.edu.cn}
\affiliation{School of Physical Science and Technology, Lanzhou University, Lanzhou 730000, China}
\affiliation{Lanzhou Center for Theoretical Physics $\&$ Key Laboratory of Theoretical Physics of Gansu Province, Lanzhou University, Lanzhou 730000, China}
\affiliation{Beijing Computational Science Research Center, Beijing 100084, China}


\begin{abstract}
Parity describing the symmetry of quantum mechanics wavefunction under space inversion transformation not only plays an essential role in solving quantum systems but also can be used to manipulate and measure the motional quantum states of such hybrid quantum systems as quantum Rabi model(QRM) and/or its variants through parity measurements. Here we address an exotic parity behavior of the QRM in its superradiant phase by numerical exact diagonalization, namely, the parities of eigenstates of the QRM behave irregular in the strong coupling regime but the sum of parities for each pair of eigenstates beginning from the ground state remains vanishing. It is found that this exotic behavior originates from the comparability of the photon distribution in the odd and even components of Fock basis when the eigenenergies of each pair of eigenstates approach enough to each other and physically is due to the emergent double-well potential induced by the strong coupling between the single-mode photon field and the two-level atom. The result not only uncovers the physics not known previously in the QRM but also makes an intrinsic limitation on the measurement precision of motional quantum states through parity measurements in modern quantum science and technologies. 
\end{abstract}
\maketitle

\section{Introduction}
Parity is an intrinsic and intriguing property of wavefunction in quantum mechanics \cite{Landau1956}. It characterizes the symmetry of quantum states under space inversion transformation. If the parity operator of a system commutes with its Hamiltonian, then the eigenvalues of the parity operator read $\pm 1$, indicating even($+$) or odd($-$) parity of the corresponding quantum states, which are taken as good quantum numbers to label the quantum states. For the meanings of parity one can refer to the standard harmonic oscillator, a textbook example \cite{Landau1956}, or to the quantum Rabi model (QRM), a building block of light-matter interaction \cite{Rabi1936, Rabi1937}. Due to the robust nature of parity, it finds wide applications ranged from mathematical physics \cite{Braak2011, Braak2019} to modern quantum science and technologies such as circuit quantum acoustodynamics \cite{Chu2018, Wollack2022, vonLupke2022}, quantum metrology \cite{Bollinger1996, Leibfried2004, Goetz2018, Bin2021} as well as quantum information processing\cite{Raussendorf2007, Riste2013, Saira2014, Kelly2015, vanDam2019, Ender2022}. In particular, it has been pointed out by Royer \cite{Royer1977} that the parity operator has a direct relationship with the Wigner function \cite{Wigner1932}, which makes it possible to manipulate and measure motional quantum states of quantum systems through direct measurements of parity\cite{Chu2018, vonLupke2022}. This motivates us to further explore the nature of parity, especially in some fundamental building block models such as the QRM and/or its variants because of their importance in fundamental science \cite{Solano2011} and wide applications in hybrid quantum systems \cite{Kurizki2015}.

In this work we address an exotic parity behavior of the QRM in its superradiant phase by numerical exact diagonalization. In the previous paper \cite{Yang2022} we found that the photon population of quantum eigenstates of the QRM in the superradiant phase exhibits some interesting features due to the action of induced double-well potential. Explicitly, with increasing coupling strengths, once the superradiant phase transition happens, the photon distribution in Fock space behaves like Poissonian statistics, and then makes a transition to the statistics of Gaussian unitary ensemble, even to the statistics of Gaussian orthogonal ensemble for the exicited states. Though we know exactly the physical reason why the photon distribution behaves as such, the question why the photon distribution in a fixed state can behave like the level distance statistics of many states \cite{Bohigas1984} remains to be understood. Here we would like to present another unusual phenomenon for the parity of eigenstates of the QRM in the superradiant phase with more stronger coupling strengths than those previously explored \cite{Yang2022}. As usual, if the parity operator of the QRM commutes with its Hamiltonian, the parity of eigenstates should remain unchanged in the whole parameter space. However, it is found surprisingly that this is not the case. In the normal phase and the near of the superradiant phase transition point, the parity indeed behaves regularly as expected. When the coupling strength is further increased to certain value dependent of the eigenstates considered, the parity of eigenstates begins to behave irregularly, seeming to be chaotic-like. More explicitly, this thing happens in the interior of the pairs composed in order of the ground state and the first excited states, of the second excited state and the third excited state, and so forth, with opposite parity in each pair. The reason for the chaotic-like parity is as follows. With increasing coupling strength, the eigenenergies of each pair approach to one another, and thus the corresponding eigenstates have a trend to become degenerate. However, before the eigenstates in the pair become completely degenerate, the photon populations in even and odd components of Fock basis become comparable, which makes the parities of each eigenstate pair mixed. As a consequence, the parities of the eigenstate pair become chaotic-like. Surprisingly, the sum of the parities of eigenstates in each pair remains zero, irrespective of which phase the QRM is in. These results have been obtained by numerical exact diagonalization, in which no additional approximation has been introduced aside from finite Fock basis up to machine accuracy. Thus we believe that these results are intrinsic for the QRM in the strong coupling regimes, which have not only a fundamental meaning on the physics of the QRM \cite{Belobrov1976, Milonni1983, Zaslavsky1981, Graham1984, Kuse1985, Bonci1991, Braak2011, Braak2019}, but also a profound implication on the applications that manipulate and measure motional quantum states through parity measurements \cite{Chu2018, Wollack2022, vonLupke2022} because of the intrinsic inaccuracy of parity found here.  

\section{Model and Method}
The Hamiltonian of the QRM consists of a single photon mode, a two-level atom and their coupling, denoting by $H = H_0 + H_\sigma \label{Ham0}$
\begin{eqnarray}
&& H_0 = \hbar\omega a^\dagger a,\label{H0}\\
&& H_\sigma = \frac{\Delta}{2}\sigma_x + g\sigma_z (a + a^\dagger), \label{Hsigma}
\end{eqnarray}
where correspond to the single-mode photon field and the two-level atom and their coupling with strength $g$, respectively. Here $a^\dagger (a)$ is creation (destruction) operator of the single mode photon field and $\sigma_x, \sigma_z$ are usually Pauli matrices denoting the two-level atom. For convenience, we rescale the Hamiltonian by the mode frequency $\hbar\omega$, thus the two-level interval $\Delta$ and the coupling strength $g$ used in the following are dimensionless. It is also useful to use dimensionless position-momentum operators related to the destruction (creation) operator by $a = \frac{1}{\sqrt{2}}\left(\xi + \frac{\partial}{\partial \xi}\right)$  and  $a^\dagger = \frac{1}{\sqrt{2}}\left(\xi - \frac{\partial}{\partial \xi}\right)$ to view the wavefunctions we are interested in.

\begin{figure}[h]
\begin{center}
\includegraphics[width = \columnwidth]{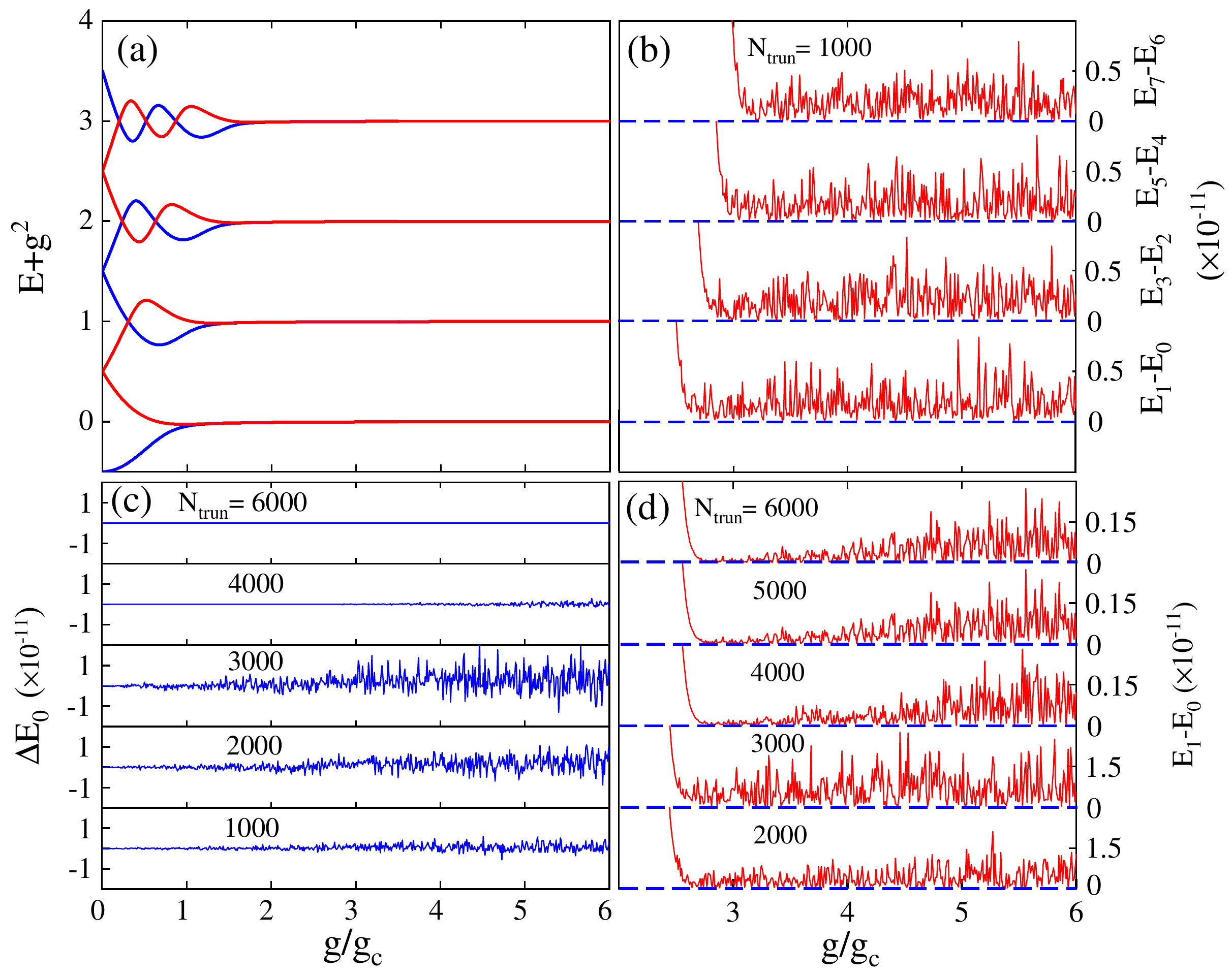}
\caption{(a) The energy levels of the ground state and the first seven low-lying excited states obtained by numerical exact diagonalization as functions of the coupling strength scaled by $g_c = \sqrt{1+\sqrt{1+\frac{\Delta^2}{16}}}$ \cite{Ying2015}. The blue and red solid lines denote odd and even parity with $\Delta = 1$. A constant of $g^2$ is added for clarity. (b) An enlarged view of the energy levels with even parity (red thin solid lines) and odd parity (blue dashed lines) in the strong coupling regimes by refering to the corresponding energy level with odd parity in (a). (c) The differences of ground state energies for different truncations of Fock basis by referring to that of $N_{\text{trun}} = 5000$. (d) The differences of eigenenergies of the first excited state (red thin solid lines) and the ground state (blue dashed lines, as reference for each truncation itself) for different truncations of Fock basis. }\label{fig1}
\end{center}
\end{figure}

We use numerical exact diagonalization method to solve the QRM. In general, it is sufficient to truncate the Fock basis up to $N_{\text{trun}} = 1000$ in the coupling strength we study below. In order to confirm this, we extend the number of Fock basis up to $N_{\text{trun}} = 6000$, and here a convergence at machine accuracy has been obtained, e.g., see Fig. \ref{fig1}(c) for $N_{\text{trun}} = 6000$, vanishing in the entire coupling regime up to $g/g_c = 6$. Since the physics we address below does not need so high accuracy, we limit ourselves to the case of $N_{\text{trun}} = 1000$. Fig. \ref{fig1}(a) shows the eigenenergies for the ground state and the first seven low-lying excited states, which are gathered as pairs beginning from the ground state. Following Refs.\cite{Braak2011, Braak2019}, we add a constant of $g^2$ to the energy levels, and thus the energy levels with odd and even parity approach to each other with increasing coupling strength, and particularly, they become almost degenerate in the strong coupling regimes. This intriguingly simple energy level structure is consistent with those obtained by the analytical ``G-function" formalism \cite{Braak2011, Braak2019}. However, the almost degenerate energy levels involve intrinsic difference, overlooked in the literature, which is the starting point of the following discussion.

\section{Results and Discussion}
Figure \ref{fig1}(b) shows the energy differences for each pair of energy levels with even and odd parity, respectively. It is observed that (i) these two seemingly degenerate levels are in fact no degenerate; (ii) oppositely, the energy levels become chaotic-like in a small but visible size of $10^{-12}$, which is at least three order higher than the machine accuracy, excluding obviously the possibility of numerical error. In order to confirm this, we show the energy differences between the first excited state and the ground state for different truncated basis in Fig. \ref{fig1}(d). It is noted that the differences in the cases of $N_{\text{trun}} = 5000$ and $6000$ are completely same; (iii) the coupling strengths starting to enter into the energy level fluctuation regimes are dependent on the excited states but the energy fluctuation amplitudes are almost the same; and (iv) most importantly, the energy fluctuations in each pair of energy levels keep always a finite amplitude, but the pairs remain intriguingly simple structure, consistent with those in Refs. \cite{Braak2011, Braak2019}.  

\begin{figure}[h]
\begin{center}
\includegraphics[width = \columnwidth]{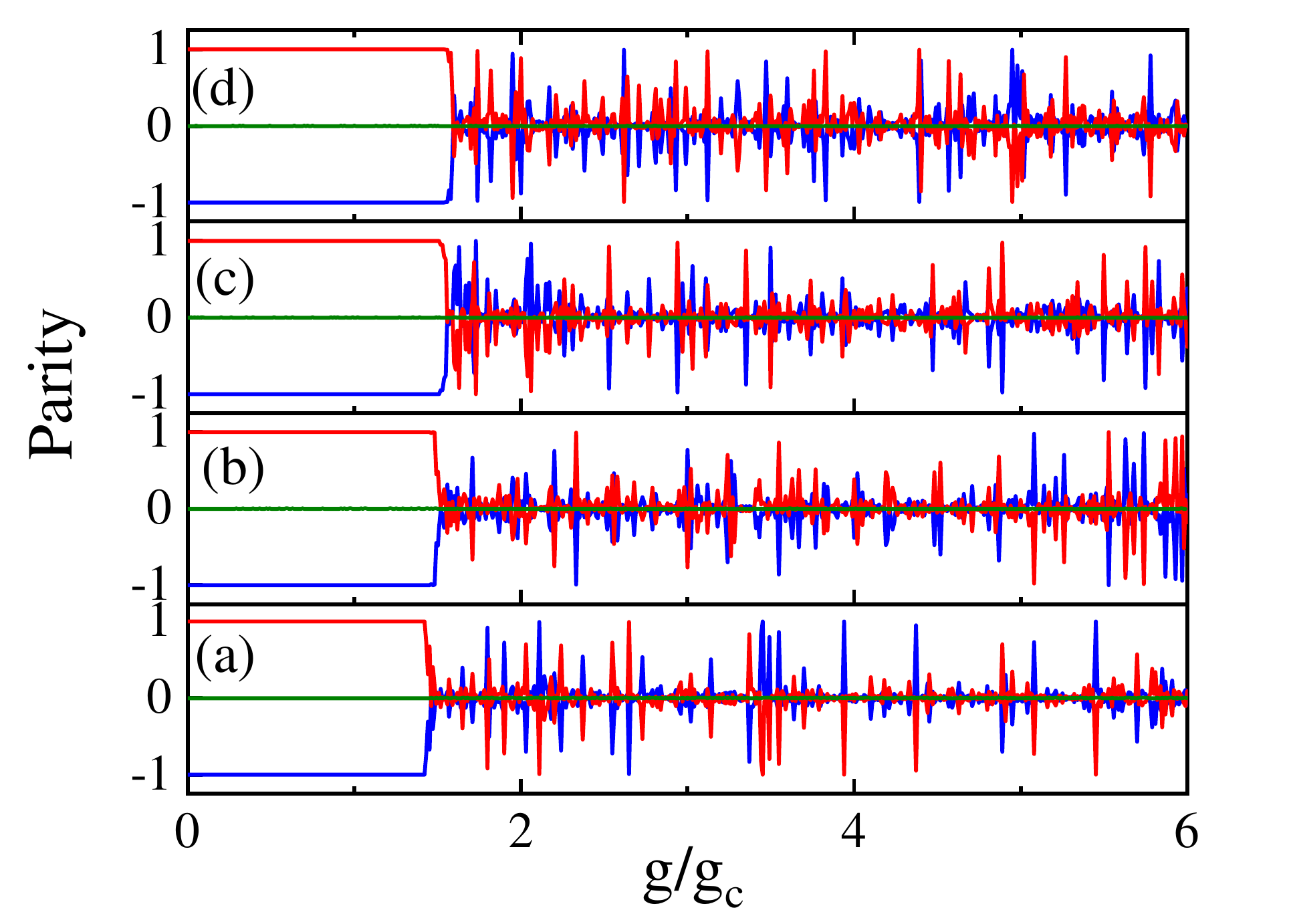}
\caption{The parity of each pair of energy levels (a) $\{E_1, E_0\}$; (b) $\{E_3, E_2\}$; (c) $\{E_5, E_4\}$; and (d) $\{E_7, E_6\}$ as a function of the coupling strength $g/g_c$. The blue lines denote odd parity and the red ones denote even parity. The green lines denote the summation of parity in each pair of energy levels. The parameter used is $\Delta = 50$. The superradiant phase transition happens around $g/g_c = 1$. The parity is not obviously broken but begins to become chaotic-like at certain coupling strengths slightly dependent of the levels of the excited states.}\label{fig2}
\end{center}
\end{figure}
The observation (iv) indicates that the QRM as a whole is still integrable in a sense that the pairs of energy levels remain ordered. However, in the interior of each pair of energy levels, it looks irregular in the strong coupling regimes, named chaotic-like behavior in the following. Besides the small energy fluctuations, one wonders what becomes irregular. It is easy to check that the parity operator of the QRM $P = \sigma_x e^{i\pi a^\dagger a}$ commutes with the Hamiltonian, namely $[H, P] = 0$, which means that the parity is a conserved quantity. However, the commutation relation is general and should be satisfied in all parameter regimes, which indicates that the parity begins to fluctuate in the strong coupling regime in order to satisfy with the commutation relation with the Hamiltonian. Indeed, the parity for each pair of energy levels is chaotic-like, as shown in Fig. \ref{fig2}, where the average of the parity $\langle P\rangle$ of the ground state and first seven low-lying excited states is given. Here the blue lines denote the odd parity and the red ones denote the even ones and the green lines are the summation of parity of each pair (see caption of Fig. \ref{fig2}). In addition, in order to show more sharply the relationship between phase transition and parity, we take the parameter $\Delta = 50$. As pointed out previously \cite{Yang2022}, the QRM enters into the superradiant phase around the coupling strength $g/g_c = 1$, at which the parity symmetry is not broken. However, the parity in the interior of each pair of energy levels begins to behave chaotic-like at certain coupling strengths, consistent with the points at which energy levels begin to fluctuate. Interestingly, for each pair the summation of the parity remains exactly to be zero, as shown by green lines. This confirms the observation in Fig.\ref{fig1} that the chaotic-like behavior emerges only in the interior of each level pair with opposite parity but there is no irregular behavior in between level pairs. 

\begin{figure}[tbp]
\begin{center}
\includegraphics[width = \columnwidth]{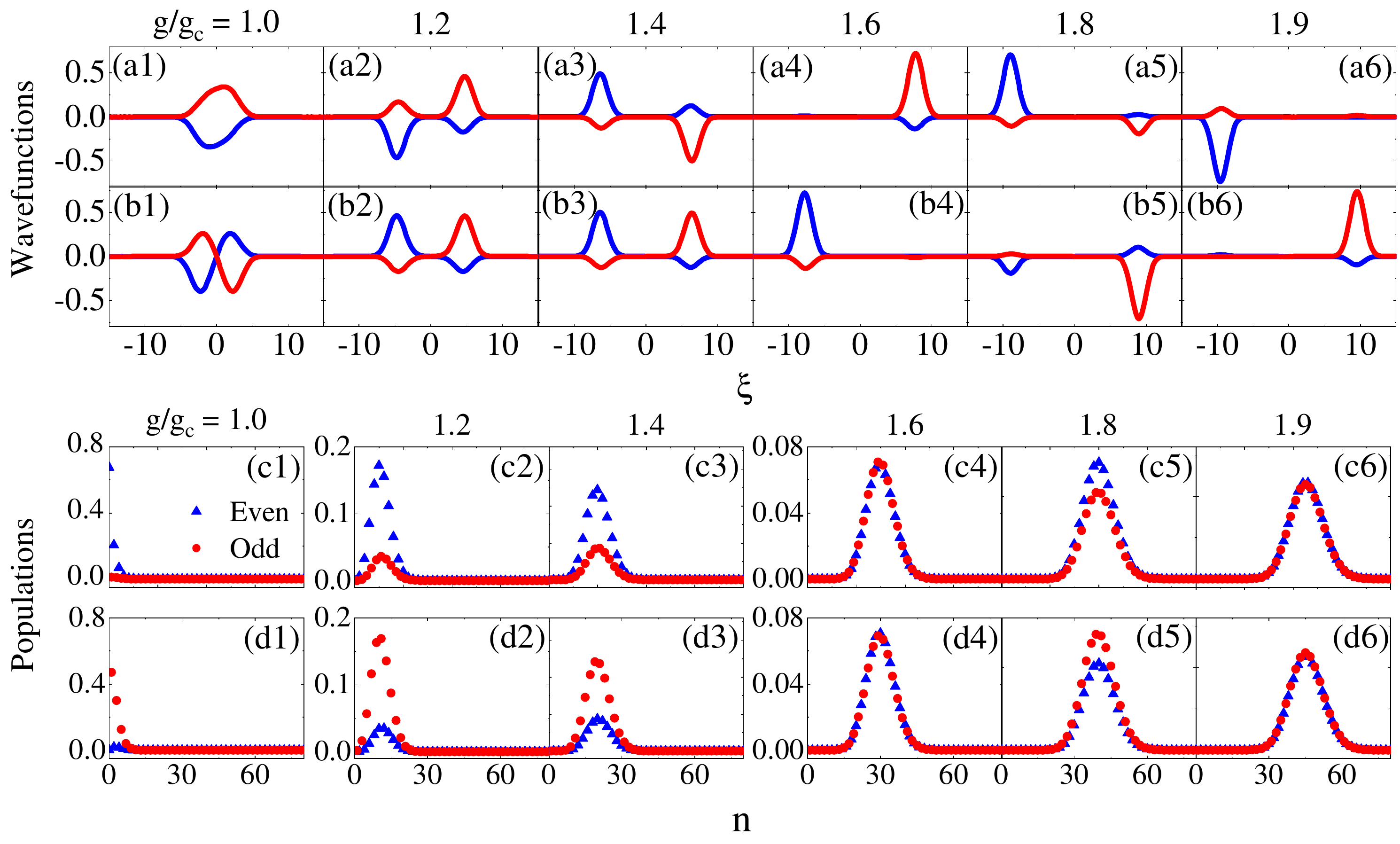}
\caption{\textbf{The upper two rows}: The wavefunctions of the ground state (a1-a6) and the first excited state (b1-b6) for six typical coupling strengths. The blue and red solid lines denote two components of the wavefunctions. \textbf{The lower two rows}: The populations of photon for the ground state (c1-c6) and the first excited state (d1-d6) for the correspoding coupling strengths. The parameter used is the same as Fig. \ref{fig2}. The blue triangles and red dots denote the even and odd components of the Fock space.}\label{fig3}
\end{center}
\end{figure}

\begin{figure}[tbp]
\begin{center}
\includegraphics[width = \columnwidth]{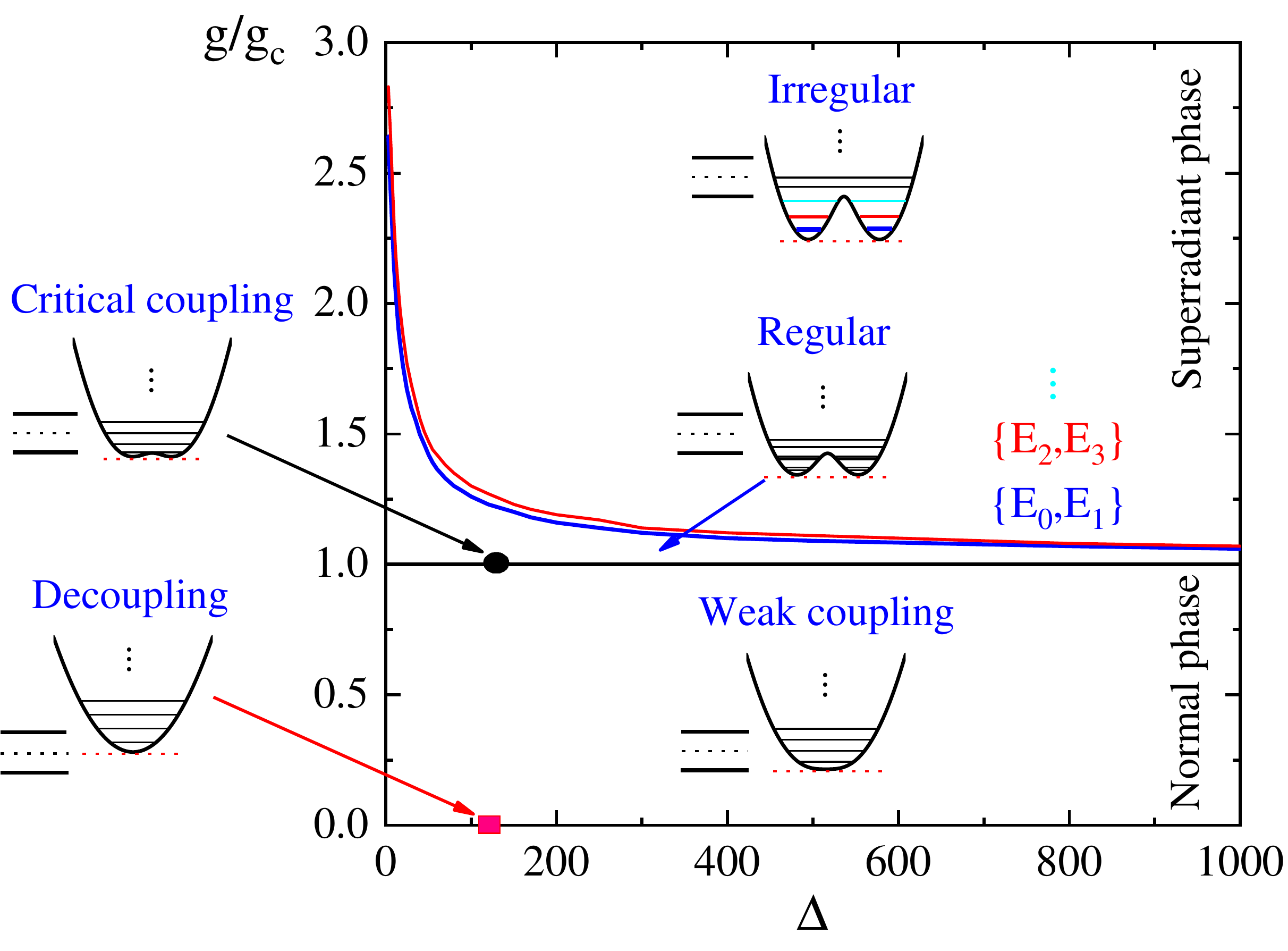}
\caption{The phase diagram of the QRM as functions of the coupling strength $g/g_c$ and the two-level interval $\Delta$ consisting of the normal phase and the superradiant phase separating roughly by black line with $g/g_c = 1$. The superradiant phase is further separated by deep-blue line or red one into the irregular and regular regimes. The irregular parity behaves chaotic-like, as mentioned in the text. Here the deep-blue and red lines denote the onset that the parities become irregular in the pair of energy levels of the ground state and the first excited state $\{E_0, E_1\}$ and of the second and the third excited states $\{E_2, E_3\}$, and so forth. The boundary further shifts upward if more higher pairs are considered such as $\{E_4, E_5\}$, and so on. In each regime, the QRM consisting of a two-level atom, the single mode denoting by a harmonic oscillator and their coupling is schematically presented with the change of an effective potential, where the black-dotted lines represent the zero point of energy and the red-dotted line the minimum of the potential. The red square represents the case without coupling, in which the two-level atom and the single mode is independent of each other and the single mode is a standard harmonic oscillator. In the weak coupling regime, the minimum of the potential declines to $-\frac{\Delta} {2}$ and the near of the minimum begins to become blunted but the minimum of the potential still locates at $\xi = 0$. Around the superradiant phase transition point, the point of $\xi = 0$ becomes a local maximum of the potential and a tiny ``Mexicoian cap" forms, which means that the superradiant phase transition begins to happen. As a result, there are two minima around $\xi = 0$. In this situation the ground state wavefunction begins to separate. Further increasing the coupling strength, an effective double-well potential forms and the minima of the potential further decline. In this regime, no obvious broken symmetry is observed and the pairs of energy levels such as $\{E_0,E_1\}$, $\{E_2, E_3\}$, and so on, begin to form, but the separation of the energy levels in the pairs is still visible. This is the non-chaotic regime. Even further increasing the coupling strength, the minima of the potential further decline and the QRM enters into the chaotic-like regime, in which the separation of the energy levels in the first few pairs of the energy levels becomes invisible and a robust doule-well potential dominates the chaotic-like behavior in this regime.}\label{fig4}
\end{center}
\end{figure}
The chaotic-like behavior can be further observed by the wavefunctions and the population of photons in Fock space for the ground state and the first excited state, as shown in Fig. \ref{fig3}. The parameter used is the same as that in Fig.\ref{fig2}. The parity of the ground state and the first excited state is indeed not broken, even up to $g/g_c = 1.4$ in the present case. Here it is noticed that the ground state wavefunction with odd parity remains antisymmetric and that of the first excited state keeps symmetric. Once the system enters into the chaotic-like regimes, the wavefunctions for both the ground state and the first excited state show non-symmetric, but it seems that they are symmetric in between, as shown in the cases of Fig.\ref{fig3} (a4)-(a6) and (b4)-(b6). Associated with this observation, it is interesting to check what happens the photon population. When the system does not enter into the chaotic-like regimes, the photon population of the ground state in even Fock states is higher than that in odd Fock states and the situation for the first excited state is just opposite. This is understandable because of the parity symmetry of the states. However, the photon populations in the odd and even Fock basis for both states become comparable, as shown in the lower two rows in Fig. \ref{fig3}, when the system enter into the chaotic-like regimes. Obviously, the fact of comparable photon populations in odd and even Fock basis leads to the missing of parity symmetry in each eigenstate. However, it is quite interest to notice that the symmetry between the eigenstates in each pair still remains, which means intrinsic integrability of the QRM as a whole. 

The phase diagram shown in Fig. \ref{fig4} summarizes our main results. As functions of the coupling strength and the two-level interval, the phase of the QRM is divided into two regimes, one in the normal phase in the weak coupling regimes and the other is the superradiant phase in the strong coupling regimes. The boundary is marked roughtly by a horizontal black line given by $g/g_c = 1$. In the normal phase, the QRM is integrable in the sense that the parity symmetry is kept \cite{Braak2011, Braak2019}, as shown in Fig. \ref{fig2}. On the contrary, the superradiant phase in strong coupling regimes show more interesting properties, which is further divided into other two regimes, one is the regular regime with conserved parity symmetry and the other is irregular regime with chaotic-like parity, which has not been explored in the literature. This is the main finding of the present work. 

It is in order to check what is the physics behind. It is known that with increasing of the coupling strength, the two-level mix with the single mode more tightly, and as a consequence, a superradiant phase transition occurs. Associated with this phase transition, an effective double-well potential forms, wihch is reminiscent of the Landau phase transition theory \cite{Landau1980}. In the case that the coupling strength is not so strong, the separation between the energy levels in a pair is still visible, and the interplay between these two approaching energy levels is still weak, thus the parity symmetry of these two states keep unchanged. When the coupling strength becomes more and more stronger, the barrier between the induced double wells becomes more and more sharper, as a consequence, the tunneling between these two wells weakens rapidly. At the same time, the eigenenergies of the pairs of eigenstates approach enough to one another. Under this condition the only interplay between these two approaching eigenstates is the transfer of the photon populations in the odd and even Fock basis, as shown in the lower two rows in Fig. \ref{fig3}, resulting in the broken parity symmetry for each state. Interestingly, the transfer of the photons or the fluctuation of the photon populations happens locally only in the interior of each pair of eigenstates, thus  the summation of the parity in each pair of eigenstates remains exactly to be zero.

Some remarks are in order. It is well-known that the classical or semi-classical limit of the QRM is non-integrable \cite{Belobrov1976, Milonni1983, Zaslavsky1981}, which indicates that the QRM itself has some unusual features but a conclusive statement about the integrability of the QRM was not reached by the conventional level statistics method \cite{Kuse1985, Braak2019}. On the contrary, the integrability of the QRM was addressed as considering the solvability of the QRM due to the conserved parity symmetry \cite{Braak2011, Chen2012}. However, the wavefunction obtained in the present and the previous works \cite{Yang2022} contain more information on the QRM, having not been explored in the literature, which uncover many intriguing features including detailed behaviors of the photon populations in Fock space \cite{Yang2022} and the chaotic-like parity found here. These features have not been reported in the literature because it is not easy to obtain exactly the wavefunction for existing methods \cite{Wolf2013} or the degeneracy of the energy levels in the strong coupling regimes has been assumed \cite{Braak2019} when the tiny difference of energy levels has been overlooked. It is necessary to check it by related methods. 

As also pointed out above, the interest features found here originate from the induced double-well potential in the strong coupling regimes of the QRM. It is well-known that the double-well potential or its extensions in the classic cases \cite{Berglund1997, Bies2001, Coullet2002, Creagh2002, Igarashi2006, Su2017} and the quantum case \cite{Berkovits1997} can be chaotic, which should have an implication on the understanding of the present irregular parity and the photon populations, even the double wells found here are induced due to the strong light-matter couplings. In addition, the QRM studied here is completely quantized in nature. What is the classical or semi-classical correspondence of the found irregular parity deserves further investigation because the quantum classical correspondence is still a long-standing problem since the early period of the foundation of quantum theory \cite{Nielsen2013, Ehrenfest1927}.

\section{Conclusion and Perspective}
By numerical exact diagonalization we find that the parity of the eigenstates of the QRM behaves irregularly in the strong coupling regimes, which has not reported in the literature. This exotic behavior of parity only happens in the interior of each pair of ordered nearest neighbor energy levels. Associated with this the wavefunctions and the photon populations show typical chaotic-like signatures. Thus a phase diagram consisting of the normal phase and the superradiant phase, and the superradiant phase consisting of the regime with regular parity and that with chaotic-like parity is obtained. 

Despite the simplicity of the QRM, its rich physics already known and further uncovered here makes possible to take the QRM as starting point to investigate the more broad physics ranged from the fundamental one to the applied one. For example, the induced double-well potential resembles the popular Landau phase transition theory \cite{Landau1980} and the quantum Duffing oscillator or its variants in the nonlinear systems \cite{Lin1990, Berkovits1997, Berglund1997, Percival1998, Bies2001, Coullet2002, Creagh2002, Igarashi2006, Su2017, Chen2021PRA}. Interestingly, the formation of the effective double-well potential is here dynamical in nature, which is thus related to the dynamical tunnelling, popular in quantum chemistry \cite{Davis1981, Keshavamurthy2011} and other quantum systems such as quantum kicked top \cite{Haake1987, Chaudhury2009, Steck2009, Neill2016, Santhanam2022}. On the application side, the QRM and/or its variants can be realized by the present techniques such as ion trap \cite{Lv2018, Cai2021, Monroe2021, Mei2022}, NMR \cite{Chen2021} as well as various hybrid quantum devices \cite{Xiang2013, Kurizki2015, Zhu2019, Zhang2021}. Therefore, the present results could be tested by possible experimental observations, in particular, the direct parity measurements \cite{Chu2018, Wollack2022, vonLupke2022}. On the opposite side, the intrinsically irregular nature of the parity in the strong coupling regimes may have an implication on the limitation of the measurement precision of the parity and thus influence the manipulation and measurement of the motional quantum states in modern quantum science and technologies. 

\section{Acknowledgments}
The authors acknowledge Liang Huang and Qing-Hu Chen for valuable discussions. The work is partly supported by the programs for NSFC of China (Grant No. 11834005, Grant No. 12047501).




%

\end{document}